# MICA : Medical Intelligent Conversational Agent[*]

## How to optimize medical teleconsultations for sports patients via a conversational agent?

Laurent Cervoni[1], Tonie Fares[1] and Mehdi Roudesli[2]

[1] Talan Research and Innovation Centre, Paris, France
[2] Centre de l'appareil locomoteur de l'Estuaire, Le Havre
`tonie.fares@talan.com`
`laurent.cervoni@talan.com`
`m.roudesli@skeewai.com`

**Abstract.** The coronavirus crisis has increased medical teleconsultations (in France) by 10 times between 2019 and the first 3 months of 2020. However, the consultation and teleconsultation process have remained much the same. To improve the patient pathway and the efficiency of a session, we have developed a conversational assistant to optimize the time spent with the patient, the quality of the information transmitted to the physicians and thus the accuracy of diagnosis. One of the chatbot's interests is to give the physicians at the beginning of the consultation key elements for further inquiry. In this way, the system aims to help the patients concentrate and well describe their problem, which leads to more meaningful interactions with physicians and helps to enable them to operate more efficiently.

## 1    Introduction

Initiated in 2014 as part of the "Establishment of accommodation for dependent old persons" and confirmed for all CPAM insured in September 2018, online health did not take place until early 2020. Thus, over a one-year period (Sept. 2018-Sept. 2019), only 60,000 teleconsultations were made (compared to 500,000 expected) [1].

But things have changed due to coronavirus. In the week of March 23-29, 2020 alone, CPAM recorded 486,368 teleconsultations [2].

Teleconsultations in France have therefore increased from 1000 to 100,000 per day due to the Covid-19 epidemic, the leading player in remote medicine. Doctolib, has claimed 885,000 teleconsultations since the beginning of March, to exceed one million since January 2019.

---

[*] Work mainly carried out between 2020 and 2021

However, the virtual consultation process might differ from an actual consultation. Aiming to make teleconsultations as "real" as possible (close to physical consultations), The French High Health Authority has put in place a fact sheet explaining the good practices for the quality and safety of teleconsultation acts, that the medical professionals should follow [3]. Therefore, in this paper, we consider studies on the durations and sequences of consultations to be valid for teleconsultations as well.

With the digitization process one could hope for an improvement of practices. However, things have remained the same. **In aiming to improve the patient journey and the effectiveness of a session**, we examined the approaches that help improve the conduct of the teleconsultation.

Our research focuses on the implementation of a conversational agent MICA: Medical Intelligent Conversational Assistant seeking to help the patients concentrate and well describe their problem, which leads to more meaningful interactions with physicians and helps to enable them to operate more efficiently.

One of the objectives is to give the doctor, at the beginning of the consultation, key elements that he can deepen more quickly, but also to improve both the patient and the practitioner's perception of the "service rendered" and to consider expressed or implied expectations.

The Regional Institute of Sport and Health Medicine of Normandy (IRMS2) and Talan have therefore collaborated and reflected on the prospects of using a tool to help prepare for teleconsultation. As the IRMS2 patient base is wide (high-level athletes, amateur athletes, sedentary people wishing or needing to resume sporting activity for medical reasons, etc.), we have defined a coherent target in order to have a homogeneous sample.

Our approach consisted, based on this sample, in determining what should be the characteristics of a "pre-teleconsultation" tool to make the teleconsultation on the identified patient more efficient and more relevant.

By definition, a virtual assistant (vocal or textual) is a computer program that tries to ensure a conversation in the most similar way to a human.

Historically, the first conversational agent or chatbot would be Eliza, created in 1966 by Joseph Weizenbaum at MIT, a kind of psychiatrist avatar who could detect keywords and respond with pre-programmed phrases [10].

The aim of MICA is neither to prescribe nor to guide the patient, but to collect essential information for the practitioner and to prepare the patient for the consultation by focusing on the important elements of his pathology and thus offer them additional listening time.

As a result, **if the patient feels like their pathology has been better understood and that the practitioner has been more effective, the reciprocal feelings will be improved.**

Finally, since the chatbot's goal is to help the professional to operate more efficiently so that the patient feels satisfied and well treated, it must be able to synthesize, structure and prioritize the information provided to the practitioner.

The article is organized as follows. Section 2 provides an overview of the use of conversational agents as medical assistants. Section 3 focuses on our methodology, how the chatbot was implemented, the choice of the sample to study and the questions to ask. In section 4 we provide the results achieved. Section 5 concludes the paper.

## 2 Related work

Many chatbots and diagnostic aid tools using AI have been implemented and studied since 1966 (after Eliza). The aim of these studies is to provide the doctors with information enabling them to refine their diagnosis and propose the appropriate treatment.

The various studies or theses that have focused on the analysis of consultations with general practitioners concentrate on the durations and the ways in which a session is organized. According to a "Direction de la Recherche, des études, de l'évaluation et des statistiques" (DREES) study of 44,000 in-office consultations, "*conducted in 2002 with liberal general practitioners, consultations and visits last an average of 16 minutes.*" In practice 70% of consultations, last between 10 and 20 minutes [4]. The same values can be found in Dr. Hélène Stephan's thesis [5], which points to the diversity of the patient motifs in one session, trying then to establish a model for the success of the consultation process. She has shown that the "patient time" i.e. the time spent by the doctor listening to patient complaints represents only 10% of the consultation time.

These findings are confirmed in particular by Dr Lucie Lacombe [6] and appear similarly in other countries even though medical practice may differ from one country to another [1].

However, some studies conclude that beyond two motifs during the consultation, the physician's listening quality decreases, the risk of a medical error increases and the practitioner's satisfaction is reduced [9].

In [8], a study of "the determinants of multiple motif/requests in general medicine consultation"; it seems necessary for the author to either educate the patient to concentrate on a single motif for one consultation, or to offer practitioners a tool allowing them to process these multiple requests without increasing the risk of error or the duration of the consultation.

In [12], Mandy, a mobile chatbot that interacts with patients using natural language, understands patient symptoms, performs preliminary differential diagnosis and generates reports was implemented. The novelty of Mandy lies in the fact that it is not directed at precise diagnosis and prediction, but rather, Mandy simply provides a humanized interface to welcome patients and understand their needs and provides valuable information to physicians for further inquiry. In this way, the system aims to free up the time of healthcare staffs for more meaningful interactions with patients and help to enable physicians to operate more efficiently. It provides a patient-end mobile application that pro-actively collects patient narratives of illness and register background information. It is equipped with natural language processing (NLP) modules that understand patient's language, process the patient symptoms, and generate interview questions. Based on interactions during the interview, Mandy will generate a report for the doctor regarding the patient's symptoms and likely causes. Mandy also provides a doctor-end desktop application for the doctors to check their patient's records and interview reports.

Some chatbots have other purposes, such as, providing support to patients with chronic diseases or comorbidities: keeping track of their conditions, providing specific information and encouraging adherence to medication. In [13] the authors introduce a chatbot architecture for chronic patient support grounded on three pillars: scalability by means of microservices, standard data sharing models through HL7 FHIR and standard conversation modeling using AIML.

The paper of Fadhil [14] describes integrating the application of chatbot systems in telemedicine application to support elderly patients living in rural areas after hospital discharge. The bot acts as a medical assistance to support patients in their health condition and accompany them in their healthcare journey.

Our Medical Intelligent Conversational Agent (MICA) aims to help the professional to operate more efficiently, and have more meaningful interactions with their patients, so that they feel satisfied and well treated. We seek to answer the following questions:

Is it possible by implementing a chatbot to help the patient better prepare for his teleconsultation? Is the following teleconsultation (with the doctor) perceived as more relevant? Do the patients feel that they have been better understood?

For the practitioner, does such an approach facilitate the course of the consultation? Does the information collected allow a better understanding of the patient's situation?

# 3 Protocol

## 2.1 Methodological framework

The first prototype of MICA is made on Azure by LUIS. One of the motivations for choosing Azure is the prospect of using Cognitive Assistants as a second prototype in order to measure the motivational score more finely and to allow a more open dialogue.

To build the conversation, we went with Microsoft's Bot Framework technology, which is an open-source solution. This allows for better control over code and "under the hood" implementations.

We relied on serverless functions to generate and send the necessary elements (prescriptions and summaries) at the end of the conversation.

In Bot Framework, we have chosen to build the dialogs using the "Adaptive Dialogs" method. This choice was motivated to facilitate its adaptation for future MICA prototypes, whether to implement additional bricks or to switch to a conversation with fewer closed questions. In addition, "Adaptive Dialogs" allow a more natural management of conversation interruptions thanks to the detection of fine intentions using one Luis entity per dialogue. Finally, the readability of the code seemed best.

This method also allowed us to be able to implement LUIS only in scenarios where it was useful to increase its performance.

One of the key elements of MICA is the generation of a summary of the patient answers and health condition to the doctor.

The MICA assistant was deployed for consultations provided by two different doctors on 95 patients randomly divided into two groups. One test group ($P_{Mica}$) is interviewed by the chatbot, ahead of the teleconsultation with the doctor, while another group ($P_{Direct}$) follows a traditional teleconsultation.

To increase the likelihood of treating different medical topics in a single consultation, we have chosen to focus this study on prescribing physical activities and encourage the patient to specify whether he (or she) experiences severe pain or symptoms that may prevent him (or her) from resuming a normal activity and to detail them if necessary.

The questions asked by the chatbot are simple with systematic help to limit possible biases related to the socio-cultural level or the relationship with a digital tool. They are based on the self-evaluation questionnaire of J. Ricci and L. Gagnon, University of Montreal, modified by F. Laureyns and JM. Sene [15].

MICA choses the questions arbitrarily (from the Ricci questionnaire) depending on the patient's answers, so neither the practitioner nor the patient will know them in advance.

At the end of each consultation, a satisfaction questionnaire (survey) is sent to the patient and the practitioner to measure their feeling/feedback on the consultation. The questions are the same for both groups $P_{Mica}$ and $P_{direct}$.

For the doctor, the questions cover:

- Quality of service provided
- Quality of consultation
- Quality of data collection
- Confidence in technology
- Feeling of time saving

For the patient the questions are intended to understand whether they feel like they have been listened to, understood and if the proposed treatment seems personalized (the latter axis is intended to determine whether the patient believes the diagnosis is guided by AI).

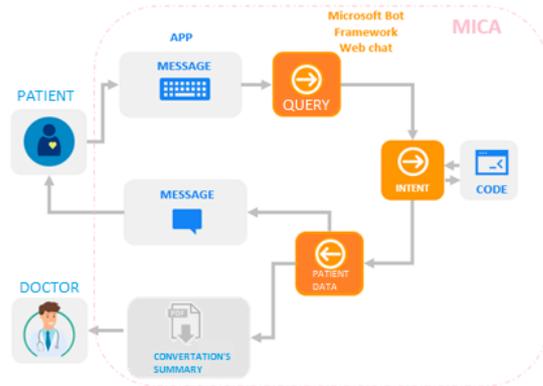

*Figure 1: Conceptual schema of the chatbot system.*

## 2.2    Data collection

The main data collected by the conversational assistant is (several **dynamic questions** for each factor)
1. Cardiovascular Risk Factors (Age and Sex are taken into account)

- Tobacco
- High blood pressure
- Hypercholesterolemia
- Diabetes
- History of infarction or stroke

2. Stress conditioning and breath or muscular limitations
3. Cardiovascular symptoms
4. Osteo-joint complaint (Mica tries to help the patient specify it if necessary)

**The concept of "multiple grounds" comes into question here, Mica "recognizes" the complaints expressed.**

5. Sport (MICA asks in detail about the patient's sports practice)

## 3    Tested hypothesis

The average age of patients in this study is 48 years. They are divided between men and women that stand in the age range between 44 and 56. All of them have already consulted a doctor at the Institute.

**Practitioners are expecting the Chatbot to help them operate more efficiently, so that the patient feels satisfied and well treated.** They expect MICA to synthesize and organize data. The information produced by the patient interview therefore highlights the key elements for the consultation:

- Patient profiling (young / sporty / sedentary with high cardiovascular risk / regular medical follow-up ...)
- Level of motivation of the subject
- Alerts for the doctor with red flags on points to be directly addressed:
  - Diabetic with non-up-to-date medical follow-up
  - Symptomatic patient
  - Many risk factors
  - …

We evaluate patient's and practitioner's satisfaction trough the questionnaires sent by the end of the teleconsultation: a score from 1 to 9 should be given to each question. We also measure the time devoted to answer the chatbot's questions by the patient and the total duration of the consultation.

We hope that, even if the consultation time is not significantly reduced, the feeling of being listened to, on one hand, and of the quality of the service rendered, on the other hand, will be improved.

## 4 Results

### 4.1 Consultation time

We observe that, the durations of consultations in the institute are in the average of what is generally observed elsewhere (between 15 and 17 minutes with longer durations for the oldest patients). Some exceptional durations were not considered because they seemed to be explained by specific pathologies requiring explanations from practitioners, outside the framework of the study of the return to sporting activity.

The use of MICA reduces the time spent in consultation by an average of 1 to 2 minutes, but this was not the primary objective.

The important point is that, in a similar time, on the practitioner's side as well as on the patient's side, the patient experience during the consultation and its efficiency are improved.

### 4.2 Relevance of interactions/ Relevance of the patient summary

The main purpose of MICA focus on improving the satisfaction of users, both doctors and patients. After analyzing the answers to the questionnaires, we noticed that: Interactions are indicated as being improved (especially among doctors who focus directly on the possible problems pre-identified by MICA).

The two populations that show the greatest satisfaction are young people and doctors. On the other hand, for the oldest, there is a slight loss of confidence linked to the presence of computers.

The time saving is insignificant during the consultation but listening skills are improved and the practitioner spends less time writing certain acts so, in practice, a few minutes of actual consultation are saved (not measured).

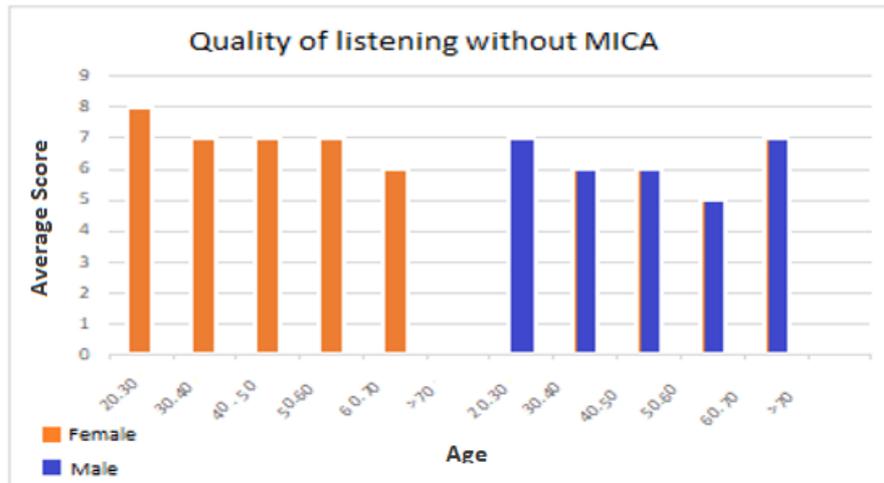

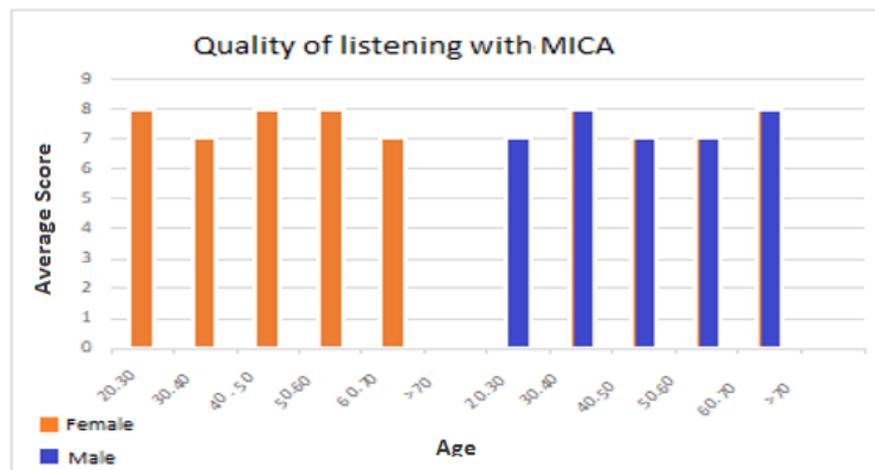

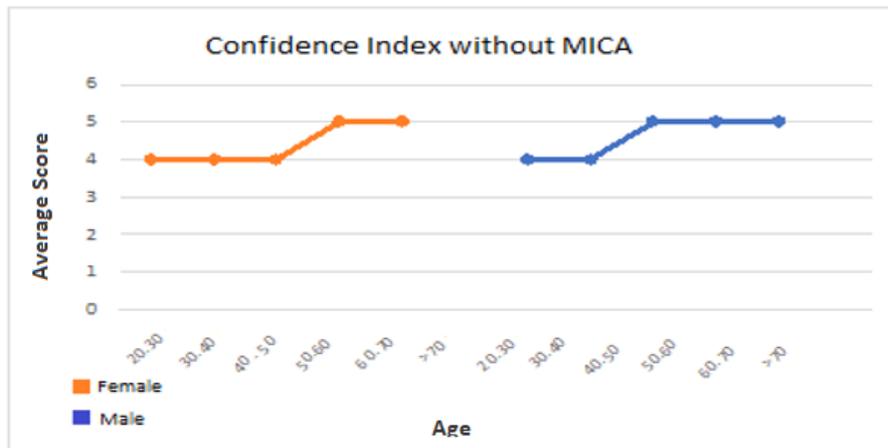

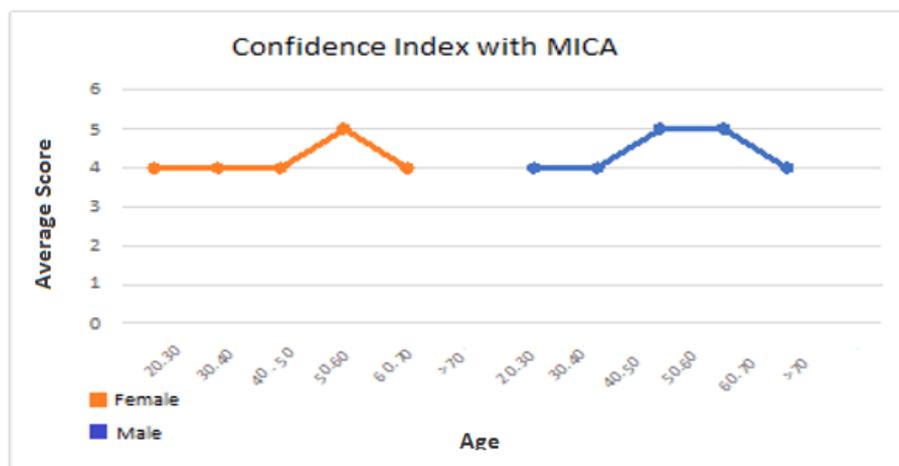

### 4.3 MICA ergonomics and user experience

The ergonomics of such a tool is a little too static. The tests used to have a dialogue tool have shown their limits because the patients go beyond the framework of the preparation of the teleconsultation by providing information that is irrelevant or too complex to manage.

if we lead patients to stay within the very specific framework of preparing a teleconsultation for a return to a sports activity program, the chatbot is very close to an "online form". Intelligence is kept to a minimum, with many questions being mandatory due to the highly targeted context.

## 5    Conclusion

The algorithm implemented in MICA remains quite simple with a tree structure of questions that adapts according to the answers provided by the patient and the generation of a summary and an automatic preparation of prescriptions.

The experience is therefore considered very interesting by doctors. However, the lack of interoperability with the business software they use is particularly restrictive:

- The link to the chatbot is sent by email with generation of an anonymized patient number
- The appointment is managed by Doctolib
- Prescriptions are managed in another tool.

Even if the functioning has been considered acceptable for a study, it presents too many constraints to guarantee data security and regular use.

The generalization of the process would require the implementation of automated security and better overall integration with practitioner software.

However, we believe that the approach is original and not very widespread for the moment in the field of medical teleconsultation. Doctors are strongly involved and very demanding for the inclusion of this type of technology in their protocols.

Since teleconsultations are asked (in France at least) to become widespread, it seems important to work to improve the care path. MICA might help achieve this goal.